\documentclass[apl,amsmath,amssymb,reprint,longbibliography,showkeys,superscriptaddress,balancelastpage]{revtex4-1}

\usepackage{textcomp}
\usepackage{graphicx}
\usepackage{dcolumn}
\usepackage{bm}
\usepackage{multirow}

\makeatletter
\def\NAT@def@citea{\def\@citea{\NAT@separator}}
\makeatother

\usepackage{color}
\usepackage{cleveref}
\crefformat{figure}{#2Fig.~#1#3}
\crefformat{equation}{#2Eq.~(#1)#3}
\crefformat{table}{#2Table~#1#3}
\crefformat{section}{#2Section~#1#3}

\begin{document}

\title[]{The dynamic evolution of swelling in nickel concentrated solid solution alloys through \emph{in~situ} property monitoring}

\author{Cody A. Dennett}
	\email{cody.dennett@inl.gov}
	\affiliation{Materials Science and Engineering Department, Idaho National Laboratory, Idaho Falls, ID 83415, USA}
	\affiliation{Department of Nuclear Science and Engineering, Massachusetts Institute of Technology, Cambridge, MA 02139, USA}
\author{Benjamin R. Dacus}
	\affiliation{Department of Nuclear Science and Engineering, Massachusetts Institute of Technology, Cambridge, MA 02139, USA}
\author{Christopher M. Barr}
	\affiliation{Center for Integrated Nanotechnologies, Sandia National Laboratories, Albuquerque, NM 87185, USA}
\author{Trevor Clark}
	\affiliation{Center for Integrated Nanotechnologies, Sandia National Laboratories, Albuquerque, NM 87185, USA}
\author{Hongbin Bei}
	\affiliation{Materials Science and Technology Division, Oak Ridge National Laboratory, Oak Ridge, TN 73830, USA}
\author{Yanwen Zhang}
	\affiliation{Materials Science and Technology Division, Oak Ridge National Laboratory, Oak Ridge, TN 73830, USA}
\author{Michael P. Short}
	\affiliation{Department of Nuclear Science and Engineering, Massachusetts Institute of Technology, Cambridge, MA 02139, USA}
\author{Khalid Hattar}
	\affiliation{Center for Integrated Nanotechnologies, Sandia National Laboratories, Albuquerque, NM 87185, USA}

\date{\today}

\begin{abstract}
Defects and microstructural features spanning the atomic level to the microscale play deterministic roles in the expressed properties of materials. Yet studies of material evolution in response to environmental stimuli most often correlate resulting performance with one dominant microstructural feature only. Here, the dynamic evolution of swelling in a series of Ni-based concentrated solid solution alloys under high-temperature irradiation exposure is observed using continuous, \emph{in~situ} measurements of thermoelastic properties in bulk specimens. Unlike traditional evaluation techniques which account only for volumetric porosity identified using electron microscopy, direct property evaluation provides an integrated response across all defect length scales. In particular, the evolution in elastic properties during swelling is found to depend significantly on the entire size spectrum of defects, from the nano- to meso-scales, some of which are not resolvable in imaging. Observed changes in thermal transport properties depend sensitively on the partitioning of electronic and lattice thermal conductivity. This emerging class of \emph{in~situ} experiments, which directly measure integrated performance in relevant conditions, provides unique insight into material dynamics otherwise unavailable using traditional methods. 
\end{abstract}


\maketitle

\section{Introduction}

The central tenant of materials science is the determining factor that structure plays in resulting properties across all classes of materials from structural alloys, to functional ceramics, to semiconductors and thermoelectrics. Most often, to make obtaining structure-property correlations tractable, one or a handful of dominant microstructural features are directly related to as-characterized material performance. Recent notable examples of such determinations include the study of short-range order on the properties of medium entropy alloys~\cite{Zhang2020}, grain size and uniformity on the performance of electrolyte membranes~\cite{Hong2021}, and point defect effects on thermoelectric and mechanical performance~\cite{Li2020}. However, in response to environmental stimuli, new microstructural features across many length scales evolve in a dynamic manner, affecting performance properties in coupled, often emergent, ways. Extreme environmental conditions which include defect-generating radiation promote some of the most dynamic microstructure evolution yet observed~\cite{Zhou2020,Nobel2020}. Nevertheless, in most circumstances, end-state performance properties resulting from exposure to coupled extreme are correlated with single-scale dominant microstructural features~\cite{Was2014}. 

This reliance on single-scale structure-property correlation is particularly acute in the study of radiation-induced swelling in structural materials, even as more advanced, chemically complex alloys are being considered for future energy applications. These complex concentrated solid solution alloys (CSAs) -- including medium and high entropy alloys -- host unique intrinsic properties including local lattice distortions and sluggish heterogeneous diffusion which have prompted their study as radiation-tolerant structural materials~\cite{Zhang2015,Zhang2019}. A growing body of experimental and computational work dedicated to nanoscale defect diffusion~\cite{Kottke2019,Zhao2017}, chemical segregation~\cite{Tuomisto2020,Cao2021}, and mesoscale microstructure evolution~\cite{Zhang2019,Fan2021} has been produced using nickel-based, single phase CSAs as a model system. Several research efforts have focused specifically on the high-damage suppression of void swelling in Ni-CSAs under ion beam irradiation. These have observed a general trend of decreased swelling with increasing chemical disorder attributed to both sluggish diffusion and preferential segregation around voids and vacancy clusters~\cite{Lu2016,Yang2017,Fan2020}. Prior efforts focused on temperatures in the range of 400--700$^\circ$C, where void swelling is expected to be strongest~\cite{Boothby2012} and have found the greatest degree of swelling to occur at about half the melting temperature in keeping with well-known trends for pure metals and traditional alloys~\cite{Fan2020}. 

While noting that the long timescale dynamical behavior of damage accumulation and swelling are of particular interest in this family of alloys, experimental studies to date have largely made use of traditional post-irradiation examination tools (electron microscopy, ion beam analysis, etc.) to capture snapshots of specific microstructure features at coarse temporal fidelity following exposure. This restriction has been driven by a combination of unique constraints involved in the investigation of irradiation-induced swelling. Significant volumetric swelling is a bulk material effect, rendering the thin, electron transparent films used in \emph{in~situ} irradiation microscopy experiments unsuitable (despite success shown in observing the earliest stages of void nucleation~\cite{Xu2013}). The few available bulk, or bulk-like, \emph{in~situ} irradiation metrology techniques available, Raman spectroscopy~\cite{Miro2016} or piezo-ultrasonic devices~\cite{Sha2020}, are largely unsuited to structural metallic samples. Recently, however, new capabilities combining laser photoacoustics with high temperature ion beam irradiation have become available which are perfectly suited to the study of the dynamic temporal evolution of swelling in Ni-CSAs where direct material property measurement provides an integrated response across all defect length scales~\cite{Dennett2019,Dennett2020a}. 

In this work, high-temperature ion beam irradiation is conducted with continuous thermoelastic property monitoring to observe the dynamic swelling response of Ni-CSAs in real-time. Single crystals of Ni and equiatomic NiFe, NiCoCr, and NiFeCoCr, as well as large-grained polycrystalline NiFeCoCrMn, are exposed to 31~MeV Ni self-ion irradiation at 550$^\circ$C to 60 displacements per atom (dpa) at the damage peak with \emph{in~situ} property monitoring. These ion beam energies are chosen to ensure that the thickness of the imposed damage layer matches the penetration depth of the surface acoustic wave (SAW) induced to monitor the real-time elastic properties. Dynamic records show elastic property evolution characteristic of face centered cubic (FCC) metals undergoing radiation-induced void swelling~\cite{Dennett2019,Dennett2020a,Dennett2018}, with notable variations between alloy chemistries. Post-exposure scanning transmission electron microscopy (STEM) confirms depth dependent swelling profiles similar to those observed previously in this family of alloys. Compared with post-exposure STEM, features of the dynamic elastic profiles indicate that nanoscale microstructural features down to the level of mono-vacancies play a significant role in the measured properties, particularly in the case of ternary NiCoCr. 

\section{Materials and methods}

\begin{figure}
\centering
\includegraphics[width=0.45\textwidth]{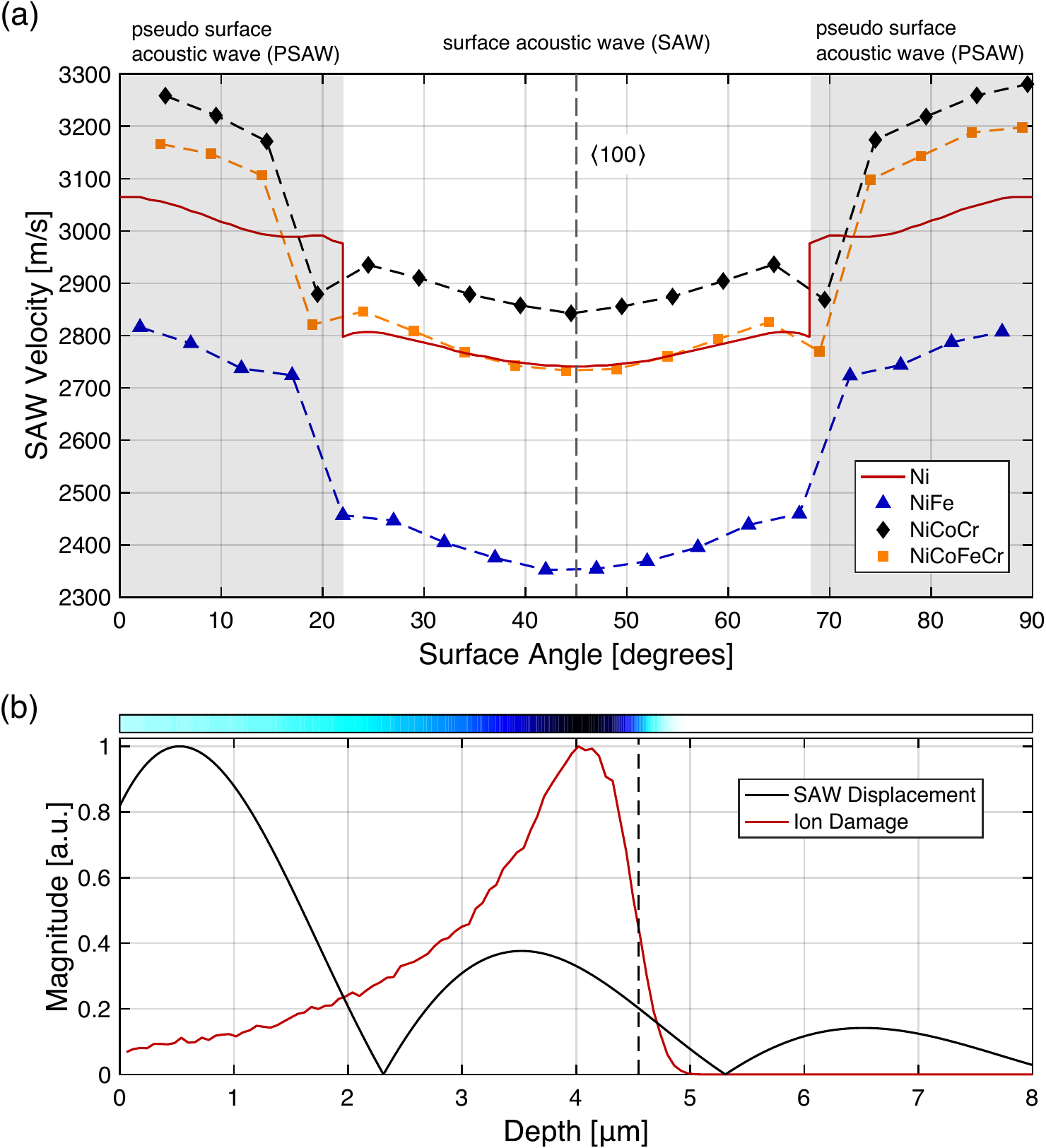}
\caption{(a) SAW velocity as a function of surface polarization on available Ni CSA \{001\} single crystals. Values for pure Ni are calculated using literature values for elastic constants and alloy samples are experimentally measured. (b) Normalized depth distribution of displacement damage induced by 31~MeV Ni$^{5+}$ ions in pure Ni at 550$^\circ$C as well as vertical SAW displacement along $\langle$100$\rangle$\{001\} showing the overlap of the damaged and excited regions. The dashed vertical line indicates the applied \emph{in~situ} transient grating wavelength of 4.54~{\textmu}m. The color map above indicates the relative level of displacement damage and is used in \cref{fig:STEM}. \label{fig:SAW_SRIM}}
\end{figure}

A single crystal nickel sample with \{001\} surface orientation, $\pm2^\circ$ misorientation, and $>$99.99\% purity was purchased from the MTI Corporation as the reference material for this study. For Ni-CSAs, ingots were prepared by arc-melting an equiatomic proportion of elemental Ni, Co, Fe, Cr, and Mn ($>$99.9\% purity), re-melting multiple times to ensure homogenity, and drop casting into copper molds. Single crystals of NiFe, NiCoCr, and NiFeCoCr were refined from arc-melted ingots through directional solidification~\cite{Bei2005}. A refining procedure for quinary alloy was not available at the time of sample synthesis. All samples were mechanically polished to a mirror finish. Prior to \emph{in~situ} irradiation, the three single crystal Ni-CSA samples were measured on a benchtop transient grating spectroscopy (TGS) experiment described previously by Dennett and Short~\cite{Dennett2017}. Briefly, TGS is a pump-probe photoacoustic technique which can be used to simultaneously capture the propagation speed of surface acoustic waves (SAWs) of various characters, depending on crystal anisotropy and surface layering, as well as surface thermal diffusivity~\cite{Hofmann2019,Dennett2018a}. The excitation takes the form of a one dimensional pulsed laser intensity profile projected onto the sample surface, generated by crossing two 532~nm laser pulses at a fixed angle with $\sim$300~ps duration, with energy of $\sim$1~{\textmu}J, and repetition rate of 1~kHz~\cite{Maznev1998}. The angle of beam crossing is selectable by the optical geometry and allows the fringe (or `grating') spacing to be selected within a range of approximately $\Lambda=1-10$~{\textmu}m. Initial characterization is carried out using a grating spacing of 4.82~{\textmu}m as determined using a single-crystal tungsten calibration standard~\cite{Dennett2017,Dennett2018a}. The resulting thermoelastic excitation is monitored through the use of a 785~nm continuous wave (CW) probe laser by diffraction from both the physical displacement associated with the induced SAW as well as any temperature-dependent changes in refractive index, amplified through the use of a heterodyne detection scheme~\cite{Maznev1998,Dennett2018a}. 

As anisotropic single crystals, the SAW velocity on \{001\} crystal faces varies as a function of angular polarization~\cite{Every2013}, with a minimum expected for a SAW along $\langle 100\rangle$\{001\} directions. SAW velocity maps on NiFe, NiCoCr, and NiFeCoCr single crystals were measured in 5$^\circ$ increments using a manual sample rotation stage to roughly identify a $\langle 100\rangle$ surface direction prior to ion irradiation experiments. \cref{fig:SAW_SRIM}(a) shows measured orientation-dependent SAW velocities for NiFe, NiCoCr, and NiFeCoCr as well as SAW velocities for pure Ni computed using an elastodynamic Green's function method~\cite{Every2013,Du2017} using literature values for the room-temperature elastic constants~\cite{Epstein1965}. A sharp jump in the measured SAW velocity at approximately 20-25$^\circ$ from the $\langle 100\rangle$ indicates the transition of the dominant acoustic mode from a purely surface-confined Rayleigh-type mode, to a pseudo-surface mode which penetrates further into the bulk of the crystal~\cite{Every2013}. For \emph{in~situ} ion irradiation experiments, the surface acoustic wave must be confined to as shallow a depth as possible to restrict the elastic propagation to the ion-modified surface layer. As such, following benchtop measurement, a $\langle 100\rangle$\{001\} direction on each single crystal was marked using a diamond scribe to ease the rough alignment of crystals prior to \emph{in~situ} measurement.

\emph{In~situ} irradiations were conducted using the \emph{in~situ} ion irradiation TGS beamline, described in detail elsewhere~\cite{Dennett2019}. This facility consists of a free-space TGS optical geometry launched into one end station of the 6~MV EN Tandem Van de Graaff-Pelletron accelerator located at Sandia National Laboratories. The minimum available transient grating spacing in this experiment is $\sim$4.5~{\textmu}m due to optical access constraints~\cite{Dennett2019}. An irradiation temperature of 550$^\circ$C was selected based on the range of temperatures for which swelling has been observed in Ni-CSA systems~\cite{Yang2019,Fan2020}. As such, a series of simulations were carried out using the Stopping Range of Ions in Matter (SRIM) code to determine that 31~MeV Ni ions best match the penetration depth of the induced acoustic wave. SRIM simulations are conducted for each alloy compositions using densities inferred from measurements of the lattice parameter at 550$^\circ$C~\cite{Jin2017} in the quick Kinchin-Pease (KP) mode using a uniform displacement energy of 40~eV for each element~\cite{Ziegler2010,Stoller2013}, accounting for the 9.5$^\circ$ incidence angle of the ion beam for I$^3$TGS targets~\cite{Dennett2019}. While recent work has shown full-cascade TRIM simulations are better able to capture the total displacement damage in multicomponent systems~\cite{Weber2019}, doses inferred from quick KP damage calculations will be referenced throughout this work to compare most directly with previous work in these systems~\cite{Lu2016,Yang2017,Fan2020}. \cref{fig:SAW_SRIM}(b) shows a normalized representative SRIM-calculated damage profile for 31~MeV Ni$^{5+}$ ions in pure Ni at 550$^\circ$C overlaid on the vertical displacement profile calculated for a Rayleigh-type SAW at this polarization~\cite{Royer1984}. Although only the profile for pure Ni is shown here, simulations carried out for each alloy composition are used in determining the applied damage in dpa. Calculated damage profiles from both quick KP and full cascade simulations are shown for each alloy composition in Supplementary Fig.~S1. Nodes in the SAW displacement magnitude indicate that portions of the surface displacement oscillate out-of-phase with one another as a function of depth from the surface. As shown, by strategically choosing the ion beam energy, the SAW will be localized to the damaged surface region, returning an elastic response representative of an average of the heterogeneously-damaged surface layer. The induced thermal wave penetration scales with $\Lambda/\pi$, indicating that for damaged microstructures the measured thermal properties will be representative of a smaller, nearer-surface volume as compared to the acoustic response~\cite{Hofmann2015,Dennett2019}.

Prior to \emph{in~situ} ion beam exposure, a surface fiducial in the form of a 1~mm square was scribed into the center of each sample to be used for laser targeting. Prior to installation, each sample was cleaned with alcohol and subjected to a 15 minute low-energy plasma cleaning to minimize any surface contamination. All high-temperature ramps and ion exposures are conducted in high vacuum better than 1.5$\times$10$^{-6}$~Torr. Once roughly aligned from benchtop polarization markings, samples were azimuthally rocked through $\sim$10$^\circ$ while the frequency of the induced SAW is monitored to ensure that they are aligned along the minimum SAW velocity direction corresponding to a $\langle 100\rangle$\{001\} surface polarization. For \emph{in~situ} TGS measurements, the excitation laser spot was $\sim$200~{\textmu}m in diameter and the probing laser spot was $\sim$100~{\textmu}m in diameter. \emph{In~situ} TGS measurements were collected using a 30~second time domain average of the 1~kHz repetition rate to increase the collected signal-to-noise ratio. In practice, measurements are collected on a 50\% duty cycle to allow for any online adjustments necessary during exposure, resulting in an ultimate temporal resolution of 60~seconds. A total applied dose of 60~dpa at the damage peak is selected for these experiments as swelling is likely to have occurred in these systems by that dose and, with an average dose rate of $1.7\times10^{-3}$~dpa/sec across all exposures (corresponding to an average ion flux of $1.9\times10^{12}$~ions/cm$^2$sec), irradiations could be completed in approximately 10 hours each. Over these long exposures, then, the 1 minute temporal resolution provides a dramatically oversampled record of the dynamic property evolution. Prior to \emph{in~situ} ion irradiation, samples were held at the target temperature of 550$^\circ$C for 3--4~hours to ensure that the entire target assembly reached thermal equilibrium and that specimens were well-annealed before irradiation. During that time, static high temperature baseline TGS measurements were collected to provide reference points for material properties. Measured values of thermal diffusivity for each alloy at 25$^\circ$C and 550$^\circ$C are shown in \cref{tab:therm} as mean values of 6--9 measurements collected for 5000 laser shots each. Values at both temperatures are in good agreement with temperature-dependent thermal transport data in these systems reported by Jin and coworkers~\cite{Jin2017}.

\begin{table}
\centering
\begin{tabular}{lcc}\hline \hline
 & Thermal Diffusivity [mm$^2$/s] & \\ \hline \hline
 & 25$^\circ$C & 550$^\circ$C \\ \hline
 Ni & $23.65\pm 0.36$ & $15.05\pm 0.05$ \\
 NiFe & $7.57\pm 0.04$ & $5.75\pm 0.07$ \\
 NiCoCr & $3.63\pm 0.14$ & $5.59\pm 0.31$ \\
 NiFeCoCr & $3.79\pm 0.14$ & $5.55\pm 0.19$ \\
 NiFeCoCrMn & $3.31\pm 0.10$ & $5.41\pm 0.12$ \\ \hline
\end{tabular}
\caption{Thermal diffusivity of the pristine CSA sample matrix at room temperature and the temperature at which \emph{in~situ} ion irradiation is carried out.\label{tab:therm}}
\end{table}

\begin{figure*}
\centering
\includegraphics[width=0.95\textwidth]{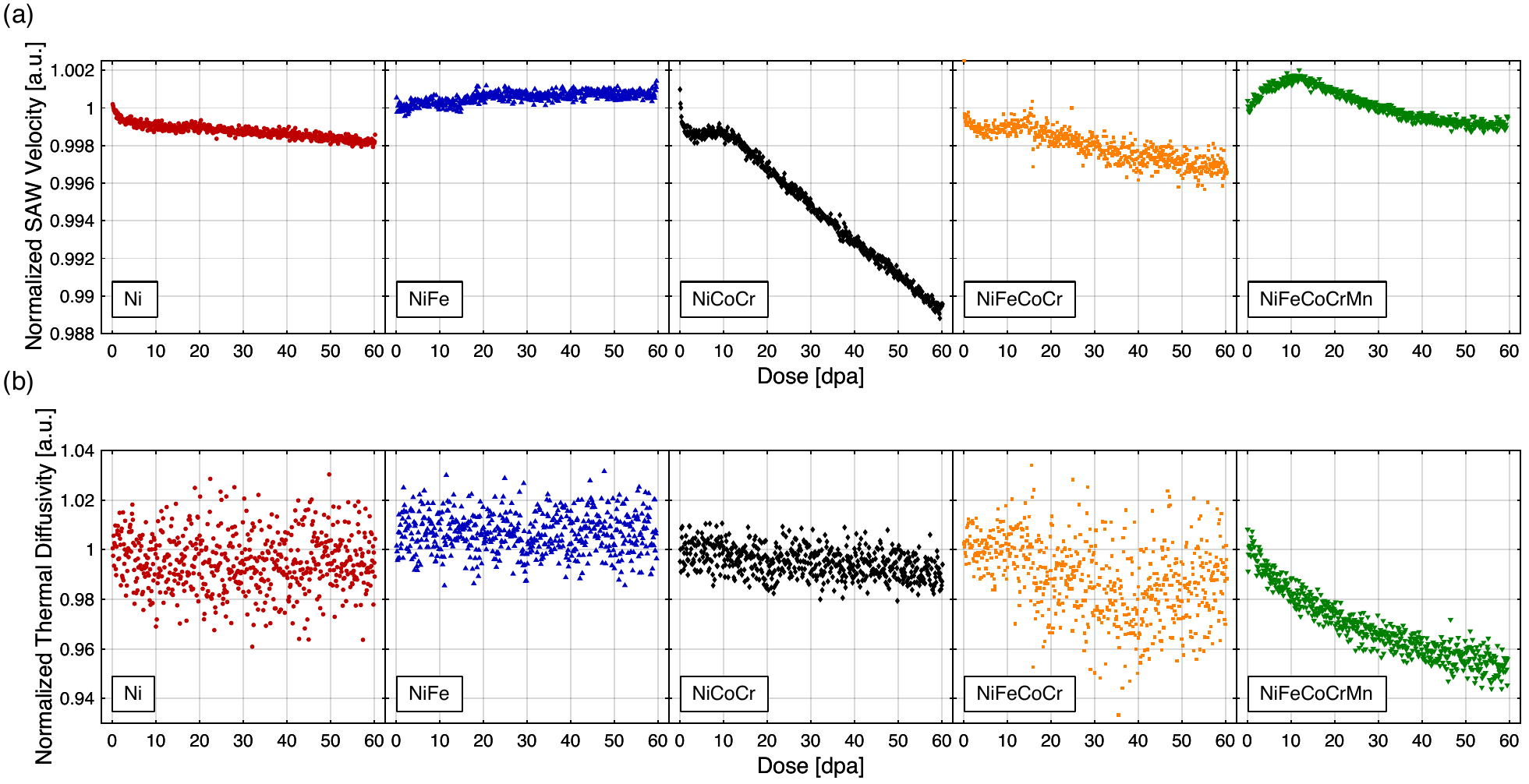}
\caption{\emph{In~situ} records of (a) SAW velocity and (b) thermal diffusivity evolution under irradiation. Values have been normalized to the first 5~minutes of exposure for each property set to make relative comparisons between different alloy compositions as the initial material properties vary by composition.\label{fig:in_situ}}
\end{figure*}

Following irradiation, cross sectional lamellae for STEM analysis were prepared using a Thermo Fisher Scientific Scios 2 Dual Beam scanning electron microscope with focused ion beam milling capability (SEM/FIB). A platinum protective layer was deposited on the surface using the electron beam at 30~keV; this layer remained intact through final thinning, which was performed with 2~keV ions. High-angle annular dark field (HAADF) STEM images were collected using an FEI Tecnai G(2) F30 S-Twin 300~kV transmission electron microscope. STEM image segmentation to calculate volumetric porosity is conducted using a procedure similar to that described in~\cite{Dennett2018}. Images are manually thresholded using ImageJ~\cite{Schneider2012} and the swelling due to large voids or cavities (hereafter ``void swelling'') calculated in 500~nm bins using the form
\begin{equation}\label{eq:swell}
S(\%)=\frac{\frac{\pi}{6}\sum_i d_i^3}{A\delta-\frac{\pi}{6}\sum_i d_i^3} \times 100,
\end{equation}
where $A$ is the image area analyzed, $\delta$ is the thickness of the lamella, and $d_i=2(A_i/\pi)^{1/2}$ are the effective diameters of each counted void determined by the void area $A_i$ through segmentation~\cite{Dennett2018,Yang2017,Fan2020}. Total void swelling within the depth measured by TGS was determined using a fixed depth of 4.54~{\textmu}m and void swelling for these regions was calculated over an area of 29--33~{\textmu}m$^2$. Depth-dependent void swelling was calculated in these relatively large bins due to the large void sizes observed in pure nickel. Analyzed lamella thicknesses varied from 135~nm to 295~nm as estimated using cross-sectional imaging during liftout. 

\section{Results and Discussion}

Normalized \emph{in~situ} records of SAW velocity and thermal diffusivity evolution in each Ni-CSA alloy are shown in \cref{fig:in_situ}. Values are normalized to velocities and thermal diffusivities measured during the first 5 minutes of exposure to enable relative comparisons between materials with differing elastic moduli and thermal transport. Examples of raw, time-domain TGS measurements from roughly halfway through each exposure are shown in Supplementary Fig.~S2. The continuous records of both ion flux recorded during irradiation and the surface-measured sample temperature are provided in Supplementary Fig.~S3. 

The dynamic evolution in SAW velocity shown in \cref{fig:in_situ}(a) is observed to vary significantly between the alloy compositions studied here. Relevant expectations for this evolution can be drawn from the prior work of Dennett and coworkers in model FCC systems undergoing void swelling through both \emph{ex situ}, post-irradiation analysis~\cite{Dennett2018} and \emph{in~situ} irradiation~\cite{Dennett2019,Dennett2020a}. In these conditions, an initial buildup of small interstitial clusters (and their interaction with any native dislocation network) is expected to increase the elastic moduli, thereby increasing the observed SAW velocity. In model systems, this initial increase is then shown to transition to a void-dominated regime where the induced volumetric porosity serves to decrease the effective modulus and thereby reduce the SAW velocity. In Ni-Fe containing alloys, work by Barashev and coworkers suggest similar behavior should be observed here as low-dose defect accumulation is dominated by interstitial-type loop formation~\cite{Barashev2019}. 

Variations of interstitial-driven stiffening (SAW velocity increase) and vacancy/void-driven softening (SAW velocity decrease) are observed across alloy compositions. In pure Ni, a steady decrease in SAW velocity is observed with no initial increase, indicating swelling accumulation begins immediately upon irradiation. For NiFe, a monotonic increase is observed, indicating that across these conditions the response remains dominated by interstitial-type cluster formation. In NiCoCr, a rapid initial drop likely associated with transient vacancy supersaturation buildup is following by a stiffening period, and finally a dramatic reduction period where a void-dominated microstructure has developed. The behavior in NiFeCoCr is similar to that of NiCoCr, albeit with a weaker relative response. Data for NiFeCoCr show more scatter in both SAW velocity and thermal diffusivity than other materials due to the reduced surface quality of this sample compared to others in the matrix. This results in a reduced single-to-noise ratio as shown in Supplementary Fig.~S2. Finally, for NiFeCoCrMn, a direct entry to a stiffening regime is then followed by the turnover expected as a result of void build-up. 

Previous investigations by Dennett and coworkers suggest that a rough determination of the ``incubation dose'' for void swelling could be estimated by the transition from interstitial-dominated stiffening to void-dominated softening behavior~\cite{Dennett2018,Dennett2019,Dennett2020a}. By linearly fitting the `rising' and `falling' segments of the SAW evolution immediately surrounding the stiffening-softening transition, as estimate for the incubation dose for each alloy composition may be made. An example of this segment analysis is shown in Supplementary Fig.~S4 for NiFeCoCrMn and the resulting incubation doses are listed in \cref{tab:incubate} for each alloy except NiFe, which by this acoustic metric does not exhibit significant swelling. While the chemically complex alloys show a significantly delayed transition into swelling, pure Ni in these conditions shows no tolerance and begins swelling immediately upon exposure. Developing physics-based models with which the incubation dose may be determined from \emph{in~situ} SAW evolution is of high priority for further investigations.

\begin{table}
\centering
\begin{tabular}{lc}\hline \hline
 & Incubation Dose [dpa] \\ \hline \hline
 Ni & 0 \\
 NiFe & n/a \\
 NiCoCr & 9.3 \\
 NiFeCoCr & 12.8 \\
 NiFeCoCrMn & 11.0 \\ \hline
\end{tabular}
\caption{Estimated ``incubation dose'' from \emph{in~situ} SAW evolution. Pure Ni begins swelling immediately upon the initiation of irradiation by this acoustic metric.\label{tab:incubate}}
\end{table}

\begin{figure*}
\centering
\includegraphics[width=0.95\textwidth]{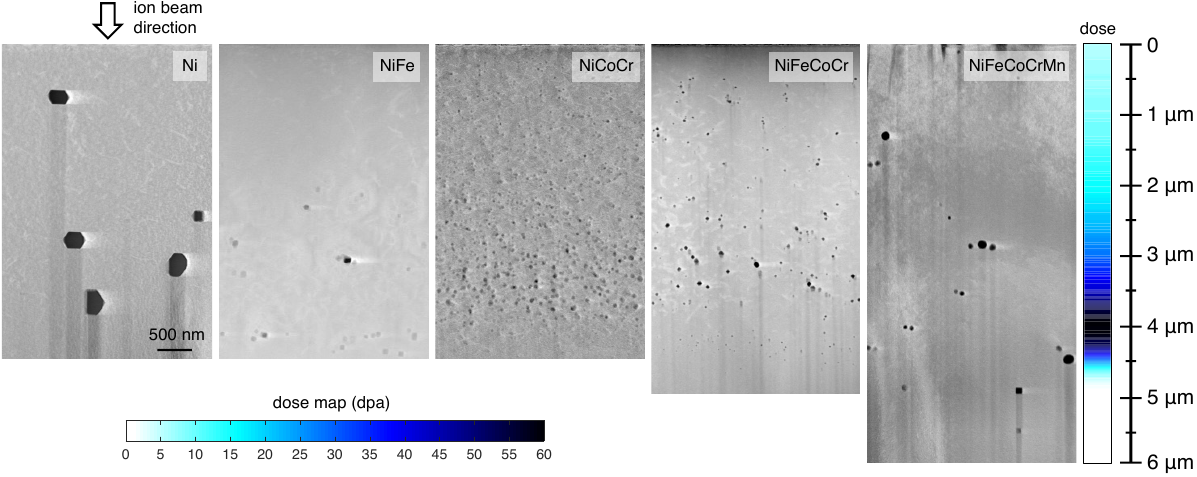}
\caption{HAADF-STEM imaging of post-irradiation microstructures for all alloys. The image depths for Ni, NiFe, and NiCoCr are equal to the transient grating wavelength. However, since large-scale defect accumulation is observed below this depth for the 4 and 5 component alloys, their micrographs have been extended to capture the entire region which experienced void swelling.\label{fig:STEM}}
\end{figure*}

In contrast to the rich behavior exhibited through changes in elastic performance, the thermal diffusivity as shown in \cref{fig:in_situ}(b) is observed to change relatively little under these conditions with a notable exception. Ni and NiFe show no discernible change in thermal performance under these conditions. Data for NiFeCoCr suffer from low signal-to-noise and no conclusions can be drawn from these data. NiFeCoCrMn, in contrast, shows a notable reduction in thermal transport in these conditions up to a total reduction of $\sim$5\% at 60~dpa. A similar, much weaker, trend of decreasing thermal diffusivity is suggested for NiCoCr, however the scatter in the data precludes any firm determination of the total relative change. In pure metallic systems dominated by electron thermal transport, little change is expected in the high-temperature thermal diffusivity in the presence of large-scale lattice defects. Several factors contribute to this expectation including weak scattering of electrons from decreased electron-phonon coupling at high temperatures~\cite{Hofmann2015a} and the length scale of voids becoming much greater than the electron mean free path~\cite{Dennett2019}. In contrast, phonons are known to scatter strongly from a variety of lattice defects. Of the alloys studied here, the relative contribution of lattice thermal conductivity has been shown to increase continually with increasing chemical complexity, with lattice thermal conductivity contributing approximately half of the total conductivity in NiFeCoCrMn~\cite{Jin2018a,Chou2009}. In this light, our observation that NiCoCr potentially shows a weak dependence of thermal diffusivity upon irradiation and NiFeCoCrMn shows a much stronger reduction is consistent with expectations based on relative electron and phonon contributions. Another factor which could play a role in the strong reduction in NiFeCoCrMn is the relatively high mobility of Mn~\cite{Vaidya2018} and its tendency towards segregation to free surfaces~\cite{Wynblatt2019}, which may be accelerated under irradiation conditions. Similar to previous work by Yang and coworkers~\cite{Yang2019}, phase decomposition from gross chemical segregation is not observed in post-exposure STEM, although Fan and coworkers have identified chemical gradients in as-irradiated NiFeCoCr under similar conditions~\cite{Fan2021}. As such, local enrichment of Mn towards the free material surface under high-vacuum and irradiation conditions, coupled with the nearer-surface sensitivity to thermal transport in TGS measurements -- sampling a depth of 1.4~{\textmu}m as compared to the 4.5~{\textmu}m sampled by the SAW -- may play a role in the observed degradation in thermal conductivity compared to the initially equiatomic lattice.  

\begin{figure}
\centering
\includegraphics[width=0.375\textwidth]{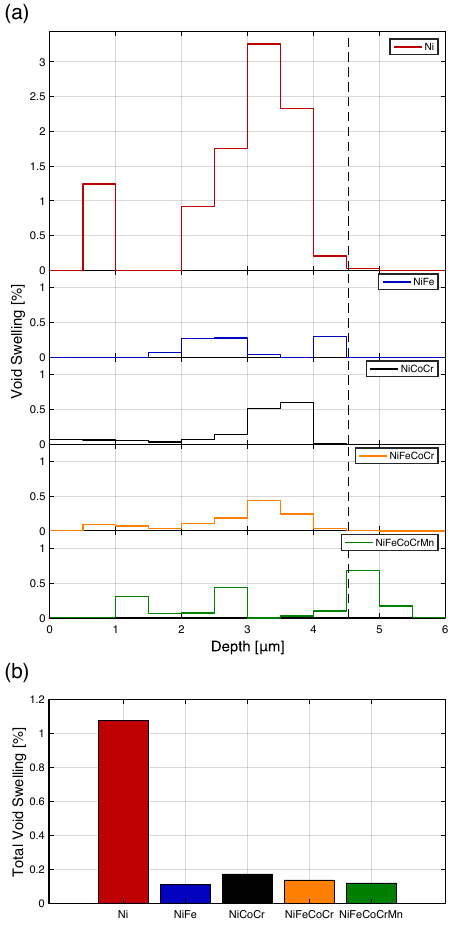}
\caption{(a) Depth distribution and (b) total void swelling measured in each alloy post-irradiation. Total void swelling values are averaged only over the TGS interrogation region defined by the grating wavelength and indicated in (a) as a dashed vertical line.\label{fig:swelling}}
\end{figure}

HAADF-STEM micrographs of post-irradiation microstructure for all five alloy compositions are shown in \cref{fig:STEM}. For Ni, NiFe, and NiCoCr, the length scale of the resulting microstructure is found to be contained within the 4.5~{\textmu}m surface layer measured using \emph{in~situ} TGS. For NiFeCoCr and NiFeCoCrMn, this affected depth is found to extend beyond that of the direct applied displacement damage from the ion beam. This behavior has been repeatedly noted in these CSA systems~\cite{Fan2020,Barashev2019,Yang2017}, the most distinct example of which observed here is the presence of a fine layer of voids in NiFeCoCr just below the peak damage location at approximately 4~{\textmu}m. Fan and coworkers have postulated that these heterogeneous dynamics are mediated by the relative concentration of species which enrich at voids (Ni and Co) and those which deplete (Fe and Cr)~\cite{Fan2020}. The relative concentration of these species changing between NiCoCr and NiFeCoCr likely plays a role in their distinct end-state microstructures. Other features, such as the separated layer of larger voids near the end of the ion range in NiFe, are consistent with previous work from Yang and coworkers~\cite{Yang2017}. Segmentation of HAADF images following \cref{eq:swell} is used to produce the depth-dependent void swelling in these materials as shown in \cref{fig:swelling}(a). In addition, the total swelling for the strictly the 4.5~{\textmu}m surface region responsible for the observed SAW response is calculated in \cref{fig:swelling}(b). For all alloys except NiFeCoCrMn, this 4.5~{\textmu}m region captures the majority of the observed void density. Voids in pure Ni are characteristic of those expected in pure metals, namely large voids hundreds of nanometers in diameter exhibiting sharp facets defined by local surface energies~\cite{Dennett2018,Yang2017}. In these conditions, voids in pure Ni have an average effective diameter of $275\pm90$~nm and are seen throughout the ion-damaged region.

General differences including a finer distribution of voids in more chemically disordered alloys and the relative ordering of total volume swelling, particularly with the performance of binary NiFe on par with that of the quaternary and quinary alloys, are also consistent with previous works where swelling tolerance is not solely determined by the number of constitute elements~\cite{Yang2017,Jin2016b}. The well-dispersed voids observed in NiFeCoCrMn which are slightly larger than those in alloys with fewer constituents are consistent with those observed by Yang et al.~\cite{Yang2019}. Despite these similarities in general trends with prior works using 3~MeV Ni ions to impose lattice damage, the total volume swelling for irradiations with 31~MeV Ni ions here is notably lower. For example, in NiFeCoCr, recent work by Fan and coworkers observed volume porosity of $\sim$1.2\% when irradiated at 580$^\circ$C to 54~dpa peak~\cite{Fan2020}, while here we observe only 0.14\% porosity when irradiated at 550$^\circ$C to 60~dpa peak. Differences in the relative contribution of electronic and nuclear stopping between low (3~MeV) and high (31~MeV) energy ions are most likely responsible for this large-scale microstructure difference. Sellami et al. have shown that energy deposited through electronic stopping of Ni ions into several Ni-based CSAs has the ability to reduce local strain and anneal pre-seeded defects from low energy ion irradiation~\cite{Sellami2019}, and ``defect healing'' through high energy ion implantation has been observed in many non-metal systems~\cite{Zhang2020a}. The electronic energy deposition as a function of depth in each alloy from both 31~MeV Ni ions (this study) and 3~MeV Ni ions (previous work~\cite{Yang2017,Yang2019,Fan2020}) is shown in Supplementary Fig.~S5, showing a large degree of extra energy deposition for 31~MeV. The synergistic effect of electronic energy deposition and defect generation in the high-energy implantations carried out here results in the dramatically reduced total swelling.

In noting the total end-state void swelling observed in each of these alloys, it is evident that the STEM-measured void swelling and the total reduction in SAW velocity measured using TGS are not directly correlated. The total volume swelling in Ni is 1.1\% while the total swelling in NiCoCr is only 0.17\%. However, the final reduction in SAW velocity in Ni is 0.2\% while that in NiCoCr is approximately 1.1\%. Features other than those observable with this magnification of STEM imaging are playing a major role in the overall dynamic response of the elastic performance. Two lower-length-scale microstructural features most likely play a role in the differences observed here. Some Ni-CSAs, notably NiCoCr, are known to undergo short range chemical ordering (SRO) upon moderate temperature annealing~\cite{Zhang2020} or irradiation~\cite{Zhang2017a}. However, recent acoustic measurement of NiCoCr in both a homogenized condition and a condition bearing significant SRO showed that the effective elastic modulus should increase from the homogenized to ordered conditions~\cite{Zhang2020}, such that any ordering present in these conditions would serve the effect of increasing the SAW velocity, opposite of the observed larger-magnitude drop. 

The most likely cause of this disproportionate reduction is therefore the retention of a significant density of nanoscale vacancy clusters and mono-vacancies below the STEM resolution limit which serve the same role as larger-scale porosity and reduce the effective elastic modulus~\cite{Dienes1952}. Prior experiments targeted towards high-temperature void swelling in these systems have shown much finer void size distributions in the ternary and quaternary alloys under investigation~\cite{Yang2017,Fan2020}, consistent with our observations in \cref{fig:STEM}. However these studies are also unable to capture nanoscale vacancy clusters which are expected are strongly influencing the modulus. A heuristic argument for the presence of these small clusters can be made from the void size distributions for each alloy shown in Supplementary Fig.~S6, where the distributions for NiCoCr and NiFeCoCr are clearly missing the tails of the size distribution towards small sizes, cut off by the minimum-detectable void diameter of 18~nm in this analysis. Further evidence for the existence of nanoscale clusters in these particular compositions is provided computationally by Wang and coworkers, who predict alloys of increasing chemical complexity should retain more smaller vacancy clusters by thermodynamic analysis~\cite{Wang2017a}. Experimental results from Tousmisto and coworkers on a similar selection of alloys up to quaternary compositions under room-temperature ion irradiation also show that increasing chemical complexity results in more small vacancy cluster retention using positron annhilaition spectroscopy~\cite{Tuomisto2020}. The weaker acoustic response observed in NiFeCoCrMn in combination with the void size distribution in Fig.~S6 suggest that nanoscale voids are contributing minimally in the quinary system, highlighting the unique role oversized Mn atoms play on kinetic defect behavior.

That mono-vacancies and clusters on the smallest scales play such a dramatic role in the material properties accessed through TGS experiments motivates a re-examination of the common definition of swelling as provided in \cref{eq:swell} and used throughout the literature on long-timescale radiation-induced evolution. This value as computed through \emph{ex~situ} microscopy analysis clearly fails to capture the the entire spectrum of defect morphologies which may be considered ``swelling.'' Indeed, the literature commonly differentiates ``lattice swelling/contraction'' at the nanoscale~\cite{Tuomisto2020}, normally captured using X-ray diffraction~\cite{Gao2019}, and ``void swelling'' at the mesoscale, the type of volume porosity captured here through electron microscopy, as separate effects. However, as evidenced through local elasticity, these different scales of vacancy-type lattice defects impose the same fundamental changes on resulting properties. As such, total swelling incubation doses identified through \emph{in~situ} TGS experiments may be more representative of the irradiation dose required to achieve breakaway behavior as effects from the entire spectrum of defect sizes and morphologies are now in fact captured. 

These integrated observations of the agglomerated role of defects of all types across length scales on bulk materials provide new insight into the dynamics of swelling and microstructure evolution under extremes beyond traditional microscopy. While this targeted demonstration has made use of complex concentrated alloys in environmental conditions of high temperature and irradiation, future applications of direct property evaluation of dynamically-evolving materials are broad, both in the classes of materials amenable to measurement using these methods and in the external stimuli compatible with the all-optical, non-contact implementation. The dynamics explored here have been interpreted in the context of a body of \emph{ex~situ} work utilizing the same methods and previous detailed microstructural characterization of this alloy system. Continued applications and extensions of these methods should be accompanied by multi-scale modeling efforts, particularly in materials and under conditions where the expected dynamic behavior is less-well understood.

\section{Conclusions}

In summary, rapidly-developing \emph{in~situ} laser-based methods allow unprecedented access to the dynamic evolution of bulk materials under extreme irradiation conditions. The thermophysical properties measured using TGS during ion irradiation capture the integrated effects of microstructure and defect evolution across length scales. Experiments carried out here on a series of Ni-based CSAs demonstrate a characteristic stiffening-to-softening transition when small-scale interstitial-type defects and loops generated at the early stages of irradiation become dominated by vacancy-type clusters and large voids, indicating a transition into a free swelling regime. Interstitial-type defect buildup in binary NiFe is sufficient to dominate the elastic response to high doses where post-irradiation STEM confirms voids have formed, while nanoscale vacancy cluster formation in ternary NiCoCr leads to a significant reduction in the measured acoustic wave velocity. Thermal transport properties are minimally affected under these conditions for alloys with low phonon contributions to thermal conductivity, Ni and NiFe, with clear thermal performance degradation observed in NiFeCoCrMn where approximately half of the total conductivity is contributed by the lattice. While the lattice contribution to conductivity of the subset of Ni-CSAs studied here increases with increasing chemical complexity, this is not generally true across materials in this class, requiring foreknowledge of thermal carrier partitioning to be able to predict radiation performance. Given both the non-homogeneous distribution of lattice damage from high-energy ions and the non-homogeneous elastic and thermal sampling imposed through TGS property evaluation, a careful consideration of all relevant length scales must be made when designing \emph{in~situ} experiments. With sensitivity to defects on the smallest scales in bulk materials, these emerging types \emph{in~situ} experiments are poised to be provide unique, length-scale-integrated insight into material dynamics under extremes.

\begin{acknowledgments}
The authors would like to thank D.L. Buller for his assistance in conducting \emph{in~situ} irradiation experiments and C.A. Hirst for helpful discussions. This work was supported through the INL Laboratory Directed Research \& Development Program under U.S. Department of Energy Idaho Operations Office Contract DE-AC07-05ID14517. This work was supported, in part, by the the DOE NNSA Stewardship Science Graduate Fellowship under cooperative agreement No. DE-NA0002135 and the MIT-SUTD International Design Center (IDC). This work was supported, in part, by the U.S. Department of Energy, Office of Nuclear Energy under DOE Idaho Operations Office Contract DE-AC07-05ID14517 as part of the Nuclear Science User Facilities. This work was performed, in part, at the Center for Integrated Nanotechnologies, an Office of Science User Facility operated for the DOE Office of Science. Sandia National Laboratories is a multimission laboratory managed and operated by National Technology \& Engineering Solutions of Sandia, LLC, a wholly owned subsidiary of Honeywell International, Inc., for the U.S. DOE's National Nuclear Security Administration under contact DE-NA-0003525. M.P.S. acknowledges funding from the US Nuclear Regulatory Commission's MIT Nuclear Education Faculty Development Program under Grant No. NRC-HQ-84-15-G-0045. H.B. and Y.Z. were supported as part of Energy Dissipation to Defect Evolution (EDDE), an Energy Frontier Research Center funded by the U.S. Department of Energy, Office of Science, Basic Energy Sciences under contract number DE-AC05-00OR22725. The views expressed in this article do not necessarily represent the views of the U.S. DOE of the United States Government.
\end{acknowledgments}

\section*{Data and materials availability}
All data are available in the manuscript and the supporting materials.

\bibliography{ref}

\end{document}


\title[]{Supplementary Figures:\\ ~\vspace{-0.75pc} \\The dynamic evolution of swelling in nickel concentrated solid solution alloys through \emph{in~situ} property monitoring}

\author{Cody A. Dennett}
	\email{cody.dennett@inl.gov}
	\affiliation{Materials Science and Engineering Department, Idaho National Laboratory, Idaho Falls, ID 83415, USA}
	\affiliation{Department of Nuclear Science and Engineering, Massachusetts Institute of Technology, Cambridge, MA 02139, USA}
\author{Benjamin R. Dacus}
	\affiliation{Department of Nuclear Science and Engineering, Massachusetts Institute of Technology, Cambridge, MA 02139, USA}
\author{Christopher M. Barr}
	\affiliation{Center for Integrated Nanotechnologies, Sandia National Laboratories, Albuquerque, NM 87185, USA}
\author{Trevor Clark}
	\affiliation{Center for Integrated Nanotechnologies, Sandia National Laboratories, Albuquerque, NM 87185, USA}
\author{Hongbin Bei}
	\affiliation{Materials Science and Technology Division, Oak Ridge National Laboratory, Oak Ridge, TN 73830, USA}
\author{Yanwen Zhang}
	\affiliation{Materials Science and Technology Division, Oak Ridge National Laboratory, Oak Ridge, TN 73830, USA}
\author{Michael P. Short}
	\affiliation{Department of Nuclear Science and Engineering, Massachusetts Institute of Technology, Cambridge, MA 02139, USA}
\author{Khalid Hattar}
	\affiliation{Center for Integrated Nanotechnologies, Sandia National Laboratories, Albuquerque, NM 87185, USA}

\begin{figure}[b]
\centering
\includegraphics[width=0.7\textwidth]{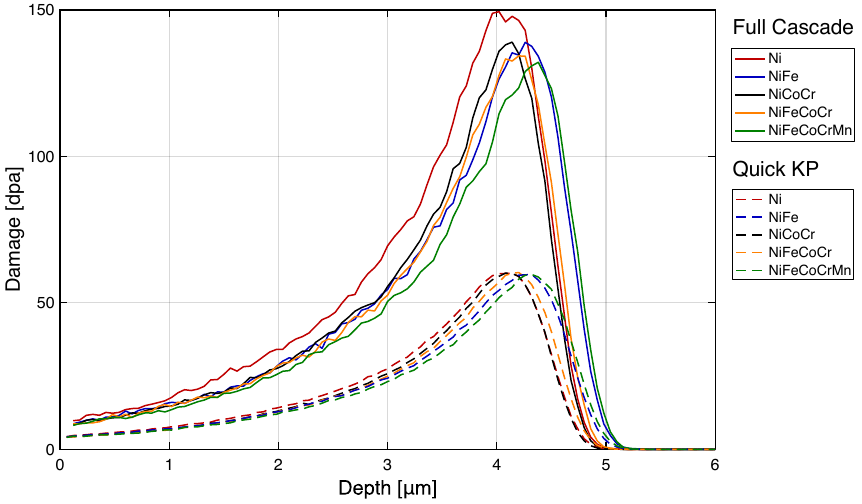}
\caption{Full cascade and quick Kinchin-Pease (KP) Stopping Range of Ions in Matter (SRIM) simulations for 31~MeV Ni ions into each of the five Ni-CSAs at 550$^\circ$C. Full cascade simulations are run for 10000 ions each and quick KP simulations are run for 100000 ions each. Dose levels in each case are scaled to the total fluence received by each individual sample which are: Ni -- $6.76\times10^{16}$~ions/cm$^2$; NiFe -- $6.79\times10^{16}$~ions/cm$^2$; NiCoCr -- $6.82\times10^{16}$~ions/cm$^2$; NiFeCoCr -- $6.84\times10^{16}$~ions/cm$^2$; and NiFeCoCrMn -- $6.85\times10^{16}$~ions/cm$^2$. Minor differences in total fluence were scaled to ensure the peak damage was equal to 60~dpa based on quick KP simulations for each case.\label{fig:SRIMs}}
\end{figure}

\maketitle

\begin{figure}
\centering
\includegraphics[width=0.55\textwidth]{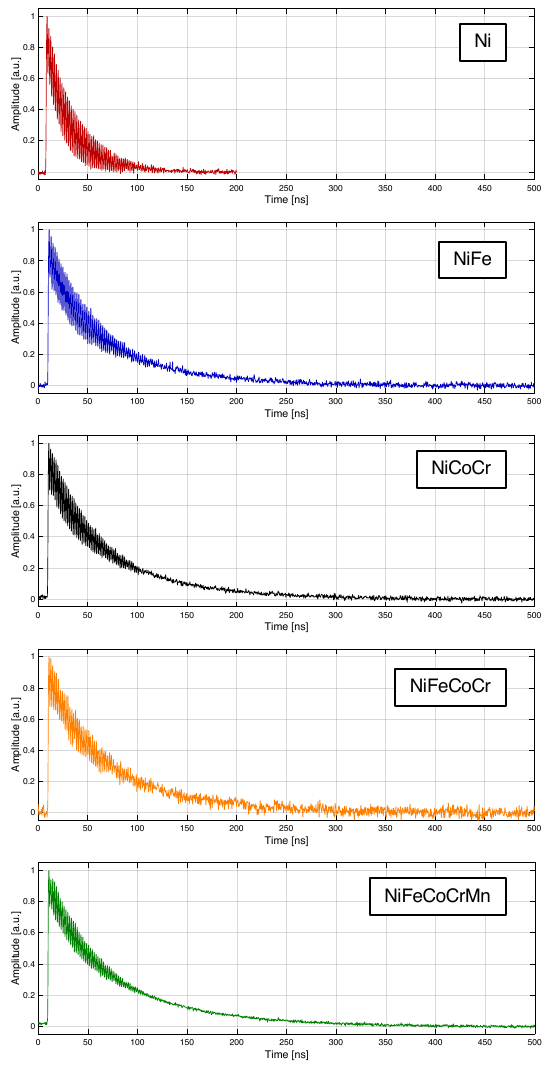}
\caption{Time-dependent TGS responses sampled from halfway through each \emph{in~situ} irradiation experiment for all five alloys. All traces represent a real-time average of 1000 excitation laser shots. The strong oscillations evident after the impulse excitation result from the Rayleigh-type surface acoustic wave and the long-timescale decay results from thermal equilibration. Alloys with higher thermal diffusivity exhibit shorter thermal decay times, allowing for a shorter collection window to be used for pure nickel (200~ns) compared to the solid solution alloys (500~ns). The higher noise level evident for NiFeCoCr is the result of lower sample surface quality, which increases diffuse laser reflection and thereby increases the scatter in the extracted values of acoustic wave velocity and thermal diffusivity.\label{fig:TG_traces}}
\end{figure}

\begin{figure}
\centering
\includegraphics[width=0.9\textwidth]{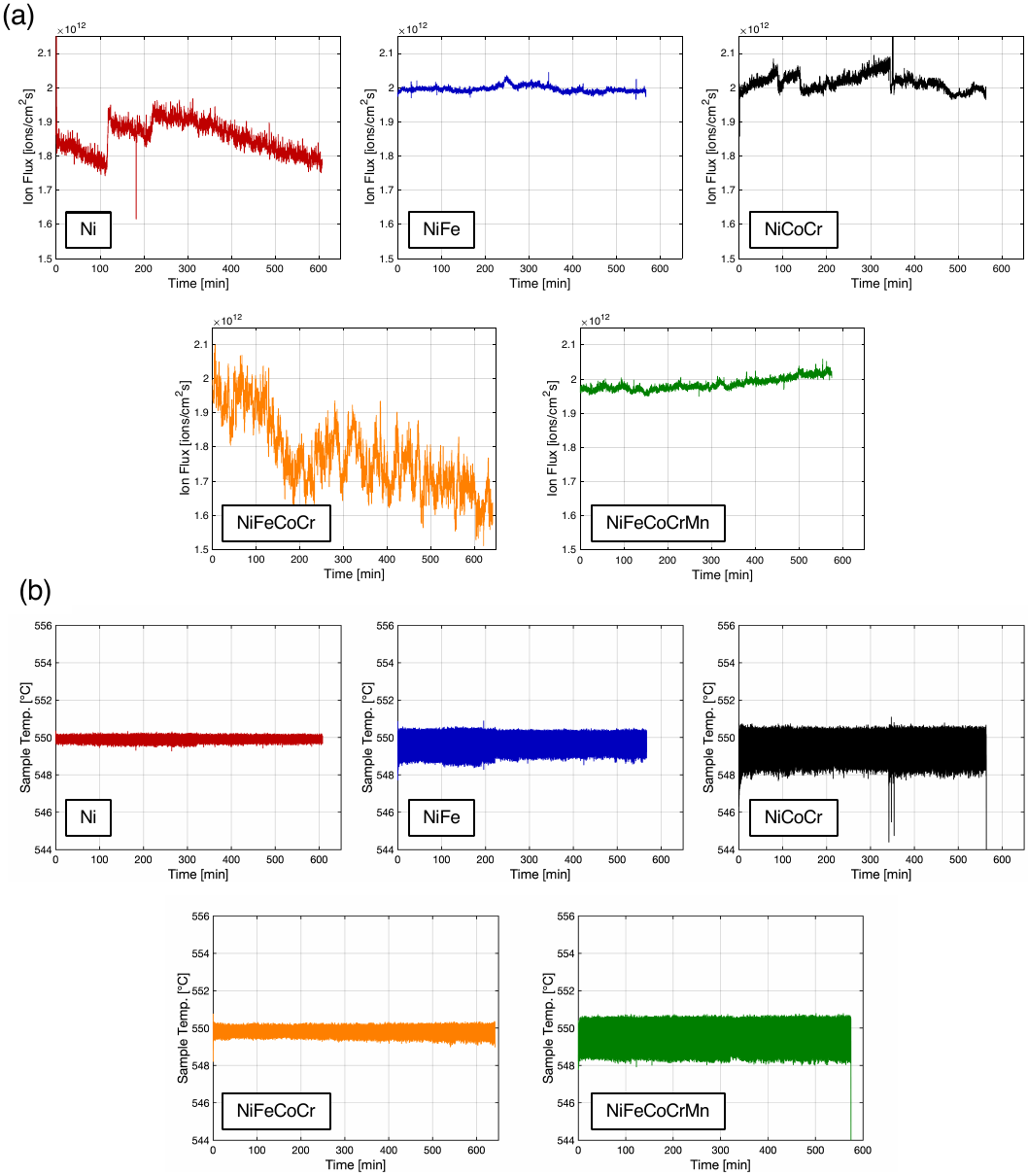}
\caption{(a) Time-dependent ion flux applied for each \emph{in~situ} irradiation experiment monitored through the use of (1) charge collection on a spinning wire beam profile monitor during irradiation to determine ion beam current and (2) optical measurement of the defoucsed ion beam spot size on an imaging quartz in the TGS sample plane prior to exposure of the target sample. Variations in ion beam current, and therefore applied flux, over time are most often to due to ion source temperature evolution or depletion. Manual control during \emph{in~situ} irradiation of the ion source is used to attempt maintain some level of uniformity in the output current, where possible. An example of this manual control can be seen in the two large steps in ion flux for the Ni case at 110 and 210 minutes. (b) Time-dependent temperature records from a surface-mounted thermocouple during each \emph{in~situ} irradiation. Heating from the applied ion beam results in some cases in an asymmetric temperate cycle around the set point at 550$^\circ$C. Sample temperatures are well-maintained through each experiment within $\pm2$$^\circ$C of the target temperature. All time-dependent profiles begin upon initial release of the ion beam onto the target sample and end after the conclusion of the final TGS collection.\label{fig:flux_temp}}
\end{figure}

\begin{figure}
\centering
\includegraphics[width=0.5\textwidth]{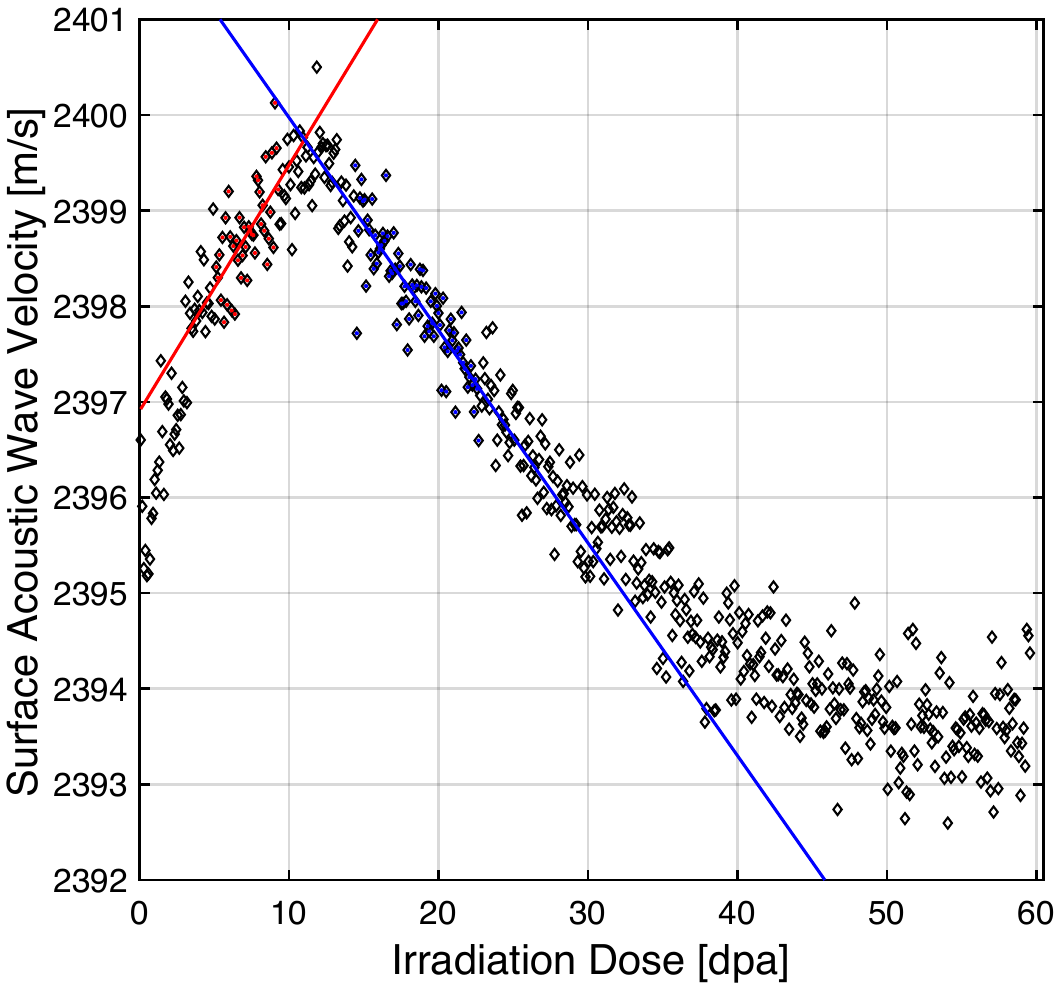}
\caption{Example segment analysis surrounding the stiffening-softening transition in NiFeCoCrMn used to estimate the ``incubation dose'' for multi-component alloys. \emph{In~situ} TGS data labeled in red is used to fit the red line for the rising segment, and data marked in blue for the falling segment. The incubation dose is defined as the intersection of these linear fits. Without a physics-based model for this transition, more complex polynomial or functional fitting to \emph{in~situ} TGS data is not currently justified.\label{fig:incubation}}
\end{figure}

\begin{figure}
\centering
\includegraphics[width=0.7\textwidth]{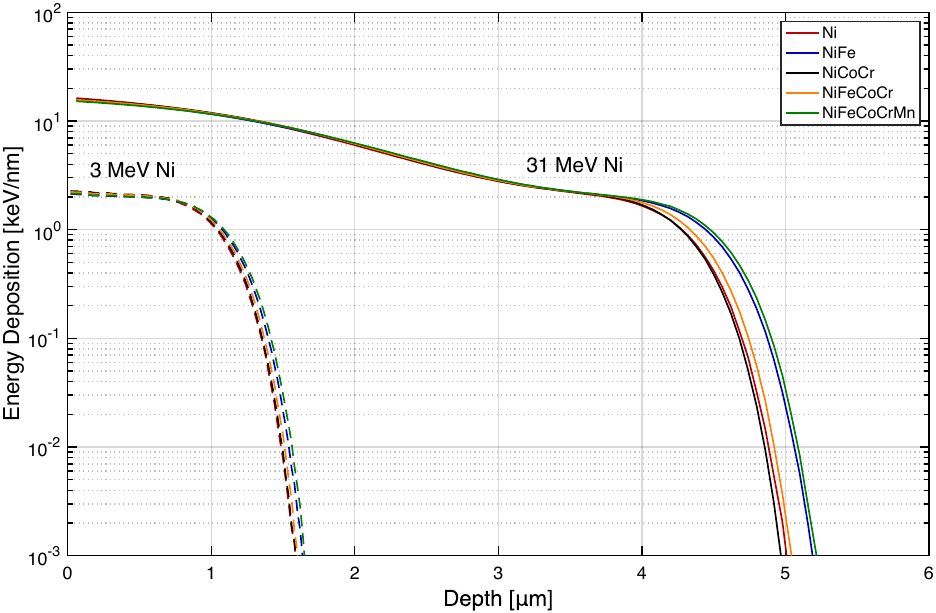}
\caption{Electronic energy deposition in each of the five alloys under consideration from both 31~MeV Ni ions (used in this study) and 3~MeV Ni ions (used in the literature) for targets at 550$^\circ$C. Energy deposition was computed using quick Kinchin-Pease SRIM calculations for 10000 ions for each conditions. As labeled, the dashed lines represent a 3~MeV ion energy and the solid lines 31~MeV. Much more electronic energy is deposited through 31~MeV ions, leading to a higher degree of ``defect healing'' during irradiation and leading to the lower overall total void swelling observed here.\label{fig:energy_dep}}
\end{figure}

\begin{figure}
\centering
\includegraphics[width=0.75\textwidth]{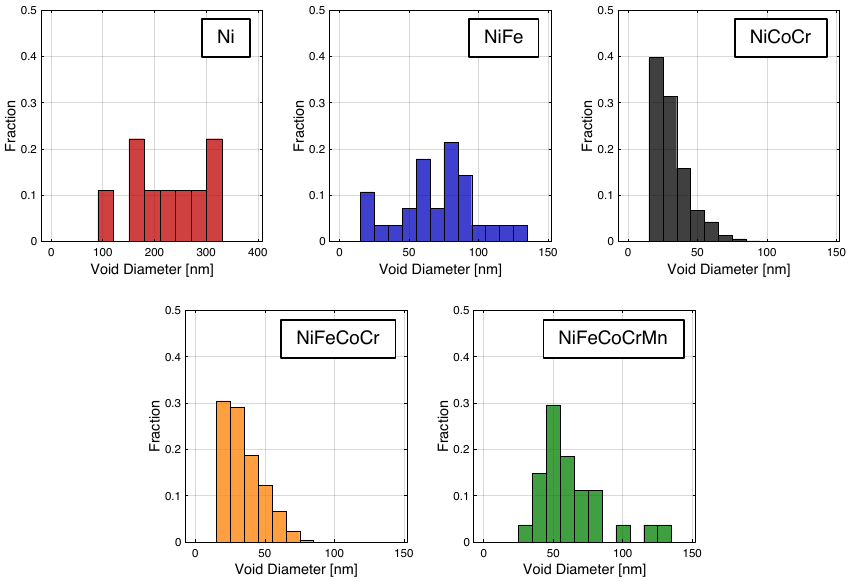}
\caption{Size distributions of voids counted in HAADF-STEM images of the post-irradiation microstructure in all five alloys as shown in Fig.~3 in the main text. The distribution for pure Ni uses 30~nm bins, while the distributions for all other alloys use 10~nm bins starting at 15~nm. The minimum void size in this image segmentation is 18~nm based on the pixel size of HAADF images at this magnification. Distributions for NiCoCr and NiFeCoCr are cut off by the minimum-resolvable void size, indicating that more small-scale vacancy clusters are likely present which are not being counted towards the total volumetric swelling, but do significantly contribute to the modulus reduction evident through \emph{in~situ} TGS responses.\label{fig:histograms}}
\end{figure}

\bibliography{ref_master}